\documentclass[sigconf,sigchi]{acmart}

\AtBeginDocument{%
  \providecommand\BibTeX{{%
    \normalfont B\kern-0.5em{\scshape i\kern-0.25em b}\kern-0.8em\TeX}}}

\setcopyright{acmcopyright}
\copyrightyear{2021}
\acmYear{2021}
\acmDOI{}

\acmConference[DaSH@KDD '21]{}{15-16 August 2021}{Virtual Conference}
\acmBooktitle{DaSH@KDD '21: 3rd Workshop on Data Science with Humans in the Loop: KDD,
  August 15-16, 2021, Virtual Conference}
\acmPrice{}
\acmISBN{}

\acmSubmissionID{7}

\begin{document}

\title{Topic-time Heatmaps for Human-in-the-loop Topic Detection and Tracking}


\author{Doug Beeferman}
\affiliation{
  \institution{Center for Constructive Communication, MIT}
  \city{Cambridge}
  \state{MA}
   \country{USA}}
\email{dougb5@mit.edu}

\author{Hang Jiang}
\affiliation{
  \institution{Center for Constructive Communication, MIT}
  \city{Cambridge}
  \state{MA}
   \country{USA}}
\email{hijan42@mit.edu}

\renewcommand{\shortauthors}{Beeferman and Jiang}

\begin{abstract}

The essential task of Topic Detection and Tracking (TDT) is
to organize a collection of news media into clusters of stories that pertain to the same 
real-world event.   To apply TDT models to practical applications such as
search engines and discovery tools, human guidance is needed to pin down the scope of
an "event" for the corpus of interest.
In this work in progress, we explore a human-in-the-loop method that helps users
iteratively fine-tune TDT algorithms so that both the algorithms and the users
themselves better understand the nature of the events.   We generate a visual overview of
the entire corpus, allowing the user to select regions of interest from the overview, and then
ask a series of questions to affirm (or reject) that the selected documents
belong to the same event.   The answers to these questions supplement the training data
for the event similarity model that underlies the system.

\end{abstract}




\maketitle

\section{Introduction}

The Topic Detection and Tracking (TDT) task \cite{allan1998topic,allan2002introduction,atefeh2015survey} 
challenges NLP researchers 
to organize a collection of news media into clusters of stories that pertain to the same 
real-world event, and to organize those events into topics.  {\it On-line} methods \cite{allan1998line,sayyadi2009event,weng2011event}, useful in alert systems, require that 
novel events are recognized and clusters are built in real-time as stories are processed, 
while {\it retrospective} methods \cite{yang1998study,allan1998topic}, useful for search and discovery tools, use all of
the accumulated data to find the best clustering.


Human feedback is vital to both kinds of systems in order to align the notion of an “event” with 
the concrete goals of a deployed application.   "Event" is defined broadly in 
the TDT guidelines -- "something (non-trivial) happening at a certain place at a certain time" \cite{yang1998study}. Recent advances in human-in-the-loop (HITL) \cite{wang2021putting} have demonstrated the effectiveness of human intervention for NLP model training \cite{kumar2019didn} and deployment \cite{hancock2019learning}. In terms of human-machine interaction, intuitive visualization tools based on topic models \cite{lee2017human,kim2019topicsifter} are developed to collect feedback from NLP non-experts. Binary or scaled user feedback \cite{simard2014ice,liu2018dialogue,kreutzer2018can} is easy to collect but sometimes oversimplify users' intentions, whereas language feedback \cite{li2016dialogue,hancock2019learning,wallace2019trick} is more informative but also challenging for machines to interpret. The feedback can be used for incremental learning \cite{smith2018closing,kumar2019didn,kim2019topicsifter} or direct manipulation of the model  \cite{li2016dialogue,kreutzer2020learning}. 

In this work we explore a human-in-the-loop (HITL) method that can
be used to fine-tune TDT algorithms so that they capture the nature of the events of interest.
In turn, the process teaches the human users about the corpus.
Our high-level approach is to give the user a visual overview of
the entire corpus, allow them to select regions of interest from the overview, and then
ask a series of questions to affirm (or reject) that the selected documents
are part of the same cluster.  The answers to these questions inform how the model, described
below, is retrained.   
Our goal is for this collaboration to enhance the efficacy of the system as both an event detector
and as a sense-making tool. 

\section{TDT systems}

Recent TDT approaches have explored both sparse and dense features. 
\citet{miranda2018multilingual}
proposes an online clustering method that represents documents with TF-IDF features, 
and demonstrate high performance on a benchmark news 
article data set.  Building on this work, \citet{staykovski2019dense} adopts a BCubed metric for 
evaluation and compares sparse TF-IDF features with dense Doc2Vec representations, showing a sizeable 
improvement on the standard data set. \citet{saravanakumar2021event} is the first to include BERT contextual 
representations for the task and achieves further improvement. Specifically, they fine-tune 
an entity-aware BERT model on an event similarity task with a triplet loss function. They 
generate triplets for each document using the batch-hard regime \cite{hermans2017defense}. In each document in a mini-batch, they mark documents with the same label as positive examples and different labels as negative examples. The hardest positive (biggest positive-anchor document distance) and negative (smallest anchor-negative document distance) examples are picked per anchor document to form a triplet. The entity-aware BERT model is trained to make the embedding distance between anchor and positive documents
closer than anchor and negative documents. Overall, this fine-tuning process 
effectively improves the contextual embedding for the overall TDT system.

Our own TDT framework (in progress) similarly mixes sparse features with dense 
features fine-tuned for event similarity. 
Instead of using hand-crafted sparse time features \cite{miranda2018multilingual,staykovski2019dense,saravanakumar2021event}, we represent the document creation time with a Date2Vec embedding \cite{kazemi2019time2vec} and infuse it with 
an entity-aware BERT embedding with the self-attention mechanism \cite{vaswani2017attention} to produce a time-sensitive dense document representation.  This captures
interactions between topic and time. 
We experiment with both online and offline triplet mining algorithms \cite{sikaroudi2020offline}
to optimally train our event similarity model.

We also add a human-in-the-loop component whereby annotators
can steer the system via an interactive visualization tool. 
Interactions with the tool are used to form triplets, which are continuously used to tune the
model's representations.  Here we aim to improve upon the process of adapting 
our system to new domains.
In this paper, we will focus on this HITL component. 

\section{Topic-time heatmaps}
\label{heatmaps}

\textbf{Motivation \& Procedure.} When we first set out to find the events in a new document
collection, we may have no labeled examples of the kinds of
events we care about.  Getting such golden data is expensive;
clustering a large set of documents into
events is time-consuming and hard to parallelize between
multiple annotators \cite{allan1998topic,allan2002introduction,atefeh2015survey}, who must coordinate on how they label events. Instead,
inspired by {\it story-story links} \cite{cieri2000multiple},
we bootstrap
our event similarity model on triplets of documents from the new data.
Such triplets can be constructed by judiciously picking
{\it pairs} of documents from the new collection, each pair annotated with 
whether it is from the same event or from different events. 
These pairwise judgments can be solicited with the guidance of annotators.
The work can be distributed across a large set of annotators without the need for them to 
coordinate on an event naming scheme; they need only 
know the guidelines for how events are to be distinguished from each other in the context
of the application.

\textbf{Interactive Heatmap.} We help annotators look for fruitful pairs in an interactive
two-dimensional heatmap visualization that positions all the documents in the corpus
by time (x-axis) and topic (y-axis).  Here
{\it time} refers to the date of the article, and
{\it topic} refers to
a projection of the event similarity model's representation
of the document text into one dimension, grouped into
{\it M} discrete buckets based on an estimate of the
number of events of interest.  
The intensity of each
cell indicates what fraction of the day's documents are mapped to that combination
of topic and date under the current model.   Each row is labeled
with the most informative words in the text of the topic it represents. 

\begin{figure}[htb]
    \centering
    \includegraphics[width=.48\textwidth]{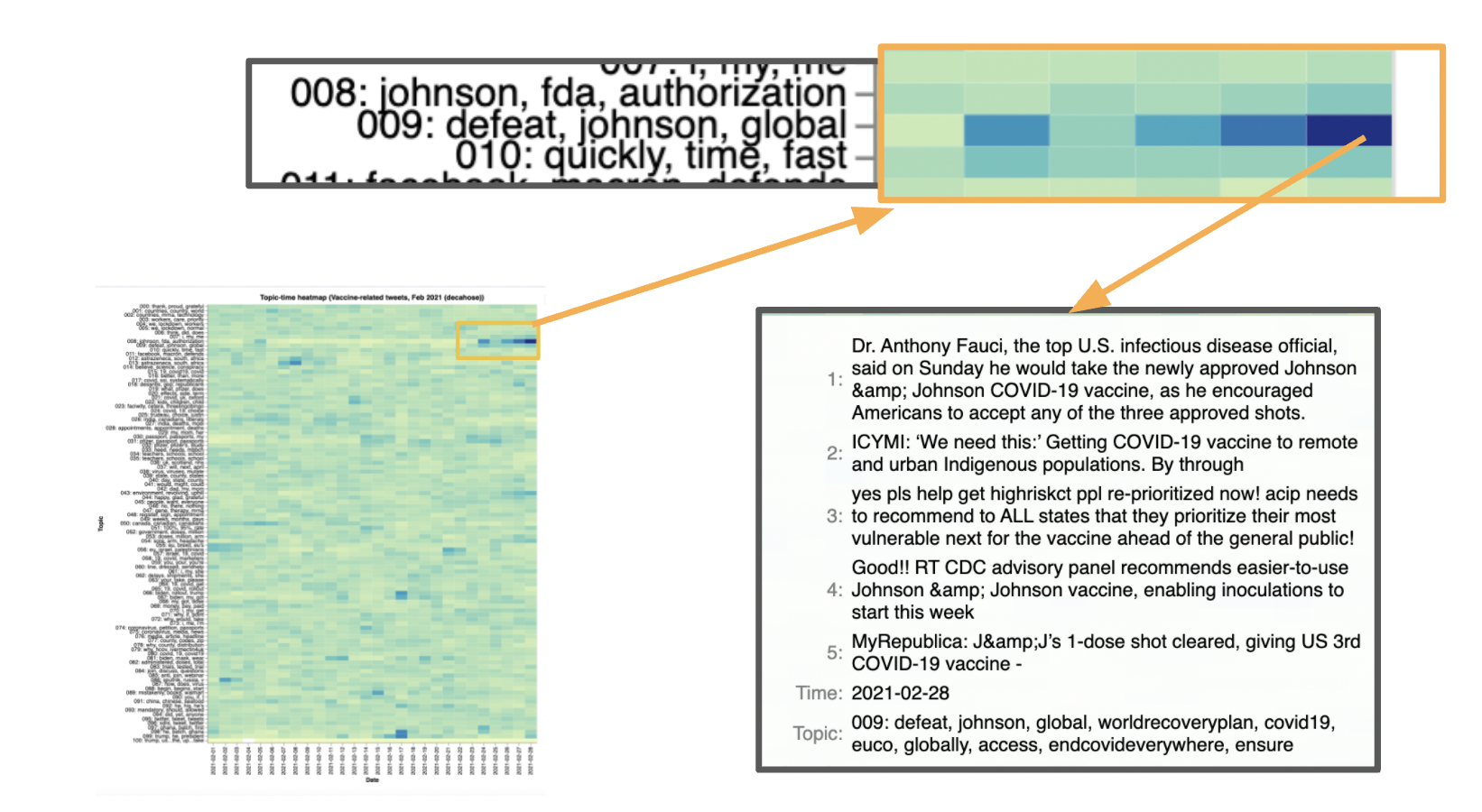}
    \caption{A topic-time heatmap (left) built from a sample of tweets from Twitter in February, 2021.  The highlighted section (top) reveals discussion of the authorization of the Johnson and Johnson vaccine in the USA at the end of that month.  Sample documents (right) are shown when the user examines one of the cells.}
    \label{fig:heatmap}
\end{figure}

\begin{figure}[htb]
    \centering
    \includegraphics[width=.48\textwidth]{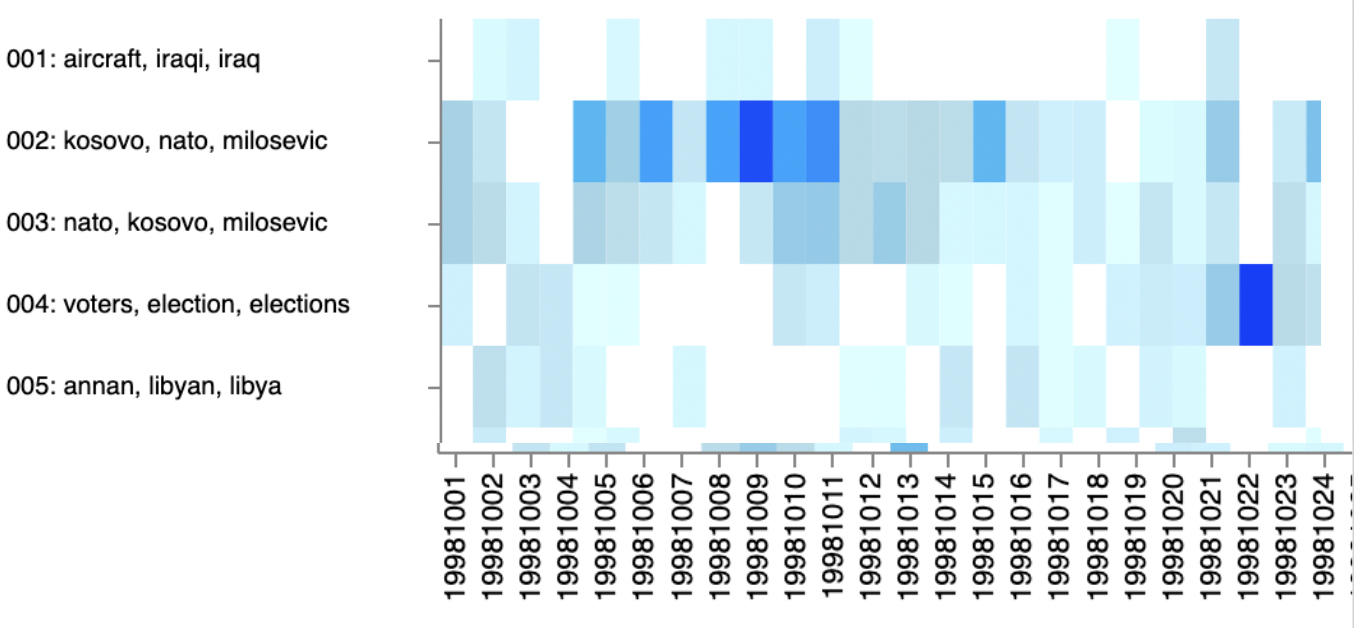}
    \caption{Part of the topic-time heatmap for the TDT3 corpus, which includes broadcast news stories starting from late 1998.  The dark cells in topics 2 and 3 reveal a burst of articles related to NATO involvement in the Kosovo War.}
    \label{fig:heatmaptdt3}
\end{figure}

Since the documents that constitute a news event typically have 
temporal and semantic locality,
stories from the same event tend to be counted in the same, or
in nearby, cells, and events often manifest as rectangular
regions.  We show heatmaps for two different 
domains, Twitter (Figure \ref{fig:heatmap}) and
broadcast news (Figure \ref{fig:heatmaptdt3}).

\textbf{User Feedback.} The annotator 
may explore the heatmap and view a sample of the documents that are counted
in each cell.
If they select a region, they are shown a randomly selected pair of
documents in the region and asked whether or not the documents belong to
the same event.  These questions are picked so as to generate useful data
for the triplet training scheme described in section \ref{heatmaps}.  
Positive pairs (those affirmed to belong to the same event) push
the representations of the even similarity model closer together for these
documents, while negative pairs (those said to be from unrelated events) 
push them apart. Triplets are constructed when positive and negative pairs share an anchor document.

\textbf{Incremental Learning.} Once a set of comparisons is collected and
the event similarity model is fine-tuned for the new detection task,
the above process can be applied iteratively;  that is, 
we can regenerate the heatmap according to the updated event similarity representations,
and solicit more feedback from the user or annotators. 
Additionally, as we now have a full TDT model
that can assign documents to event IDs, we can tabulate (and visualize)
how well the new model addresses the cumulative human feedback collected.  
A flawless event similarity model would place all of the documents
corresponding to an event into the same row.

\section{Conclusion and Future Work}

In this short paper we have outlined our in-progress work on a human-in-the-loop
topic detection and tracking system, and we have introduced a topic-time heatmap visualization that
human annotators can use to improve both the efficacy of the system on new corpora, and their
own understanding of the data.  We are in the process of measuring 
this system's efficacy compared to established techniques for collecting event
detection training data\footnote{Our software for generating heatmaps from
a corpus of timestamped documents is 
available at {\tt https://github.com/social-machines/semsearch}}. 

\bibliographystyle{ACM-Reference-Format}
\bibliography{main}

\appendix









\end{document}